\documentclass[aps,prd,reprint,superscriptaddress,nofootinbib]{revtex4-1}

\usepackage{ifxetex}
\ifxetex
   \usepackage{polyglossia}
   \setdefaultlanguage{english}
   \usepackage{fontspec}
\else
   \usepackage[utf8]{inputenc}
\fi
\usepackage{amsmath}
\usepackage{graphicx,color,overpic,mathtools}
\usepackage{amsfonts}
\usepackage{euscript}
\usepackage{mathrsfs}

\usepackage{amsthm}
\usepackage{bm}
\usepackage{amsmath}
\usepackage{amssymb}
\usepackage{setspace}
\usepackage{hyperref}
\usepackage{xcolor}
\PassOptionsToPackage{normalem}{ulem}
\usepackage{ulem}
\usepackage{tensor}
\usepackage{tikz}

\newcommand{\noprint}[1]{}

\newcommand{\tr}{{\text{tr}}}

\newcommand{\ii}{\mathrm{i}}
\renewcommand{\d}{\mathrm{d}}

\newcommand{\nn}{\nonumber}

\usepackage{braket}
\usepackage{bbm}

\usepackage{braket}
\usepackage{bbm}

\begin{document}
\title{
Relativistic causality in particle detector models:\\
Faster-than-light signalling  and ``Impossible measurements''}

\author{Jos\'e de Ram\'on}
\affiliation{Institute for Quantum Computing, University of Waterloo, Waterloo, Ontario, N2L 3G1, Canada}
\affiliation{Department of Applied Mathematics, University of Waterloo, Waterloo, Ontario, N2L 3G1, Canada}

\author{Maria Papageorgiou}
\affiliation{Institute for Quantum Computing, University of Waterloo, Waterloo, Ontario, N2L 3G1, Canada}
\affiliation{Department of Applied Mathematics, University of Waterloo, Waterloo, Ontario, N2L 3G1, Canada}
\affiliation{Division of Theoretical and Mathematical Physics, Department of Physics, University of Patras, 26504, Patras, Greece}

\author{Eduardo Mart\'in-Mart\'inez}
\affiliation{Institute for Quantum Computing, University of Waterloo, Waterloo, Ontario, N2L 3G1, Canada}
\affiliation{Department of Applied Mathematics, University of Waterloo, Waterloo, Ontario, N2L 3G1, Canada}
\affiliation{Perimeter Institute for Theoretical Physics, 31 Caroline St N, Waterloo, Ontario, N2L 2Y5, Canada}

\begin{abstract}
We analyze potential violations of causality in Unruh DeWitt-type detector models in relativistic quantum information. We proceed by first studying the relation between faster-than-light signaling and the causal factorization of the dynamics for multiple detector-field interactions.  We show in what way spatially extended non-relativistic detector models predict superluminal propagation of the field's initial conditions. We draw parallels between this characteristic of detector models, stemming from their non-relativistic dynamics, and Sorkin's  ``impossible measurements on quantum fields'' [\href{https://arxiv.org/abs/gr-qc/9302018}{arXiv:gr-qc/9302018}]. Based on these features, we discuss the validity of measurements in QFT when performed with non-relativistic particle detectors. 
\end{abstract}
\maketitle

\section{Introduction}

 Relativistic Quantum field theory (QFT) shares many of its  defining features with non-relativistic formulations of quantum mechanics. However, the two crucially depart in the sense that QFT is a theory explicitly formulated in a relativistic (possibly curved) spacetime manifold, which allows to incorporate  notions such as Einstein causality or general covariance. In particular, the formal and conceptual differences between QFT and non-relativistic quantum mechanics become patent in the context of  \textit{quantum measurement theory}, since the standard description of measurements has been found to be at odds with the notion of relativistic causality \cite{sorkin1993impossible}.

In this context, particle detector models (such as the Unruh-DeWitt (UDW) model~\cite{PhysRevD.14.870,PhysRevD.29.1047}) can be thought of as a framework to model the measurement, or `probing', of quantum fields. As their name suggests, these models were originally concocted to account for particle phenomenology in cases where a naive particle notion fails, e.g., for non-inertial motion or curved spacetimes. Generically defined to be non-relativistic quantum-mechanical systems that couple locally to a fully relativistic QFT, particle detectors have been able to deliver physical insight in scenarios ranging from fundamental problems in QFT, such as horizon radiation and the entanglement structure of quantum fields~\cite{PhysRevD.15.2738,Valentini1991,Reznik2003,PhysRevD.94.064074}, to practical descriptions of the light-matter interaction in quantum optics and quantum information experiments~\cite{PhysRevD.94.064074, PhysRevD.97.105026,PhysRevD.101.045017}.

The UDW model was first proposed by Unruh in his seminal paper \cite{PhysRevD.14.870}, with the intention of addressing the question of whether an accelerated observer will experience the vacuum of a field theory as a thermal bath. In this first work, the detector was modeled in two different, yet thought of as complementary ways. First, the detector was prescribed as a confined quantum mechanical particle in a box, and second as an auxiliary quantum field. Later, DeWitt simplified the former by further considering the detector as a point-like two-level system~\cite{DeWitts}. Yet elegant in its simplicity, the point-like model suffers from several technical complications such as spurious ultraviolet (UV) divergences when the interaction between detector and field is not switched on and off adiabatically~\cite{Louko2008}.

Beyond the point-like model, the practice of smearing the interaction of the detector with the field in spacetime has become popular in the particle detector literature. Indeed, one can consider smeared detectors as straightforward generalizations of the point-like model that do not suffer from UV divergences~\cite{Schlicht,Louko2008,Satz:2006kb}, i.e. as regularized detectors. In addition, from the practical point of view, smeared detector-field interactions 
can better capture the physics  of experimental set-ups involving, e.g, the light-matter interaction \cite{PhysRevD.97.105026,PhysRevD.101.036007,  PhysRevA.103.013703}, or the physics of superconducting qubits \cite{PhysRevA.96.052325}. Despite these advantages, smeared couplings are not devoid of their own fundamental issues: coupling non-relativistic systems to \emph{smeared} field operators causes problems with relativistic causality \cite{PhysRevD.92.104019} and the covariance of the model \cite{PhysRevD.101.045017,2020broken}.

The main subject of this paper is to deepen the analysis of the friction between \emph{smeared} detector models and relativistic causality for general detector models in curved space-times, with an emphasis in the problem of the so-called ``impossible measurements in QFT''~\cite{sorkin1993impossible,borsten2019impossible,bostelmann2020impossible}. Crucially, the causality issues we will tackle are introduced by the very fundamental construction of the model per-se, and not by extra approximations that introduce non-locality, such as the rotating-wave approximation~\cite{ PhysRevD.100.065021}, or other a-posteriori non-relativistic approximations~\cite{Papageorgiou_2019}. 

A fully relativistic measurement scheme for QFT, while technically involved, is of course devoid from causality issues (e.g., the FV-framework~\cite{fewster2020quantum}). However it is perhaps still reasonable to approach measurements in QFT from much simpler, effective, non-relativsitic detector models such as the UDW.  In this paper we will be concerned with structural aspects of UDW-type models related to the interplay between their \emph{non-relativistic} nature and  their spacetime \emph{localization} through their smearing. We will pay especial attention to the possibility of superluminal signaling in smeared models, a phenomenon that oughts to be unacceptable in relativistic physics.

This work is divided in the following sections. In section \ref{sec:Measurements} we discuss the challenges that arise in the formulation of measurement schemes in relativistic QFT, to highlight the differences with respect to the standard `quantum-mechanical' theory of measurement. This is necessary for identifying which of these challenges are relevant for detector models in QFT. In section \ref{sec:detector} we establish a unifying language in terms of which one can tackle the causality issues of general detector models in general globally hyperbolic space-times.

In section \ref{sec:faster}, we give general (non-peturbative) arguments about signaling between two  detectors, and show the absence of faster-than-light signalling, that is, the absence of signalling when the detectors' couplings  are constrained to spacelike separated regions. 

In section \ref{sec:propagation} we discuss superluminal propagation of initial data through the following set-up: The initial data is encoded in the field state through a local unitary over a region A. A detector C, partially in the causal future of region A, `mediates' as a repeater information about this initial data to a detector B that is partially in the causal future of C, but spacelike separated from region A. In this section we will establish links and parallels between this feature of detector models and Sorkin's no-go result on impossible measurements in QFT~\cite{sorkin1993impossible}. 

In section~\ref{sec:perturbation} we further analyze the interplay of the concepts of superluminal signaling and superluminal propagation in three-detector scenarios. We show that particle detector models do not suffer superluminal signalling at the first three orders in perturbation theory in the coupling constants of the detectors. We conclude and summarize in section~\ref{sec:conclusions}.

\section{ Measurement schemes in QFT}\label{sec:Measurements}

In this section we review some general considerations regarding measurements that are particular to relativity and QFT. Aspects of quantum measurement theory, as well as the measurement problem, are most commonly discussed in the context of non-relativistic quantum mechanics. This can be helpful to isolate the aspects of the problem that are present in \textit{any} quantum theory, relativistic or not. Nevertheless, when one tries to formulate the problem in the context of relativistic QFT, there is a variety of technical and conceptual challenges that come from the explicitly relativistic nature of the theory \cite{ruetsche2011interpreting}.

Despite the success of the quantum field theoretical description in high energy physics and cosmology, not much progress has been made in the development of a measurement theory in QFT (only recently in the context of the algebraic approach \cite{fewster2020quantum}).   In contrast, quantum mechanics comes with (at least a standard) measurement theory, i.e. a formalism for the mathematical description of measurements on which basis  interpretational issues can be addressed. Commonly, the physical predictions are encoded in probability distributions derived from the theory by means of projector valued measures (PVM), which assign probability distributions to sets of commuting observables following Born's rule. More generally, one can implement measurement schemes in which a system is measured with the aid of an auxiliary system, or detector, which effectively generates probabilities described by positive operator valued measures (POVM)~\cite{NielsenChuang}.  In either case, it is further assumed that quantum mechanical systems undergo a ``state update'' once a property of the system is measured, and, in non-relativistic quantum mechanics, this update is generally modelled by Luder's rule~\cite{NielsenChuang,PhysRevD.1.566}.

  These concepts of standard quantum-mechanical measurement theory happen to be incompatible with relativistic causality. Indeed, it can be proved that any localized operator valued measure (in the sense of being associated to spatial sub-intervals of a hypersurface) is incompatible with the microcausality condition, i.e. that observables associated with spacelike separated regions commute. This follows from a series of no-go theorems, the most relevant ones  being the initially formulated  by Malament \cite{malament1996defense} for PVM's and the extension to POVM's by Hegerfeld \cite{hegerfeldt2001particle}. These theorems are most commonly discussed in the context of the localisation problem and the non-existence of  position operators in relativistic set-ups, and they seem to hint the necessity of a field theoretical description of relativistic quantum physics.

Perhaps surprisingly, formulating a measurement theory that is consistent with relativity has not been any easier in the context of an explicitly relativistic quantum field theory.  The notion of instantaneous state update becomes problematic  since joint probability distributions will depend on the order in which the measurements are performed, which  generally depends on the reference frame.  A modified prescription is required for the extraction of frame-independent probabilities of successive measurements.  A covariant version of the state update rule that gives rise to frame independent probability assignments was developed by Hellwig and Kraus \cite{PhysRevD.1.566}. 
  
Moreover, in the algebraic approach to quantum field theory (AQFT), local spacetime regions are associated with local Von Neumann algebras, which should be regarded as the building blocks for a local measurement theory. However, such algebras in AQFT, under some mild assumptions \cite{fewster2019algebraic}, are known to be of type III. This type of algebras do not contain finite-rank projectors, which poses several challenges to the usual construction of measurements schemes, e.g. the Kraus representation of POVM's \cite{okamura2016measurement, ruetsche2011interpreting}. Furthermore, the type III character of the local algebras has non-trivial  consequences for the locality and the entanglement properties of the theory \cite{redhead1995more, Halvorson2004-HALEAO}.

Regarding the possibility of superluminal signaling, perhaps the most important challenge was posed by Sorkin's no-go result \cite{sorkin1993impossible} on the impossibility of (idealized) measurements in QFT.  Sorkin's work demonstrates that signaling between two spacelike separated regions A and B can be `mediated' by an operation on a third region C that is partially in the future of A and partially in the past of B. 
Since superluminal signaling is not compatible with the axioms of relativity, Sorkin's result proves that a naive `quantum-mechanical' set of measurement rules would fail in relativistic QFT. Furthermore, the  result raises the issue of the consistent description of \textit{successive} measurements when more than two measurements in different spacetime regions are involved. The issue stems from the fact that a partial causal order can be defined between pairs of extended (bounded) regions, but cannot be naturally extended to multiple regions (unless they are pointlike). Sorkin suggests that a resolution can be given in the more `space-time oriented' formulation of sum-over histories approach (See, e.g.,~\cite{anastopoulos2015measurements}).

Sorkin's result was further analyzed recently in \cite{borsten2019impossible}, where they studied (from a purely formal point of view) what conditions can be imposed as requirements for local, field-valued POVM measurements to avoid superluminal signalling. Moreover, a formal resolution of the Sorkin problem  has been proposed recently in the context of Fewster and Verch framework for measurements in algebraic QFT \cite{bostelmann2020impossible}, although a connection from the formal results to the description of experiments where quantum fields are measured remains elusive in this context.

As we discussed in our introduction, particle detector models do provide an alternative framework for the description of local measurements in QFT. Thus, we would like to study whether particle detector models present similar incompatibilities with relativity as those on Sorkin's ``impossible measurements''. Due to the non-relativistic nature of the detector system, the compatibility with the premises of the underlying relativistic theory (QFT) is generally not guaranteed. To ground the formalism of particle detectors on a relativistic QFT we need to consider the problem as  made of two related (but separate) components:  1) the possible violation of relativistic causality (related to the presence or not of FTL signalling in the framework) and 2) the possible breakdown of relativistic covariance (related to coordinate independence of predictions). In fact, these two components are separate within QFT even before any notion of measurement is introduced~\cite{earman2014relativistic,book}.  Both requirements would be easier to fulfill in the case that the detector is also modeled as a relativistic system (probe quantum field) ~\cite{fewster2020quantum}, but the main advantage of non-relativistic detectors is that they come with a standard measurement theory. For example, one can measure them projectively (in contrast to the quantum field) and infer the induced field observable associated with each outcome.

Even though the standard algebraic approach to QFT does not provide any measurement axioms, AQFT formalizes a great variety of causality conditions \cite{earman2014relativistic,RevModPhys.90.045003} that can be recruited to analyze any particular measurement model.  Typically, the microcausality axiom (i.e. commutativity of field observables in spacelike separation) is associated with non-signaling, at least when \emph{non-selective}~\footnote{A non-selective measurement is a linear CPTP map that sends a state to a convex combination of states associated with possibles outcomes of a measurement, whereas a selective measurements is a map, generally non-linear, that sends a state to one of those states associated to a particular outcome.} measurements are introduced~\cite{earman2014relativistic}. The covariance axiom guarantees the coordinate independence of the dynamics of the theory and, crucially, it is independent from the microcausality axiom. For example, we could design a theory in which superluminal signaling is not possible but the laws of physics are dependent on the reference frame in which we describe them. 

These two topics were addressed separately by one of the authors in the context of detector models in QFT. Covariance requires examining the transformation properties of the interaction Hamiltonian density between the field and the detector under changes of reference frames~\cite{2020broken}. Causality issues in Unruh-DeWitt detector models were analyzed in~\cite{PhysRevD.92.104019}. Also, a treatment of the causality porblem that shares some similarities with our approach has been explored for the particular case of accelerated point-like detectors in a recent paper by Scully et al.~\cite{PhysRevResearch.1.033115}. In what follows we will present a general analysis that will allow us to comment on the interplay between these two notions (covariance and causality of general detector models) as well as relate it with Sorkin's identification of the problem with standard measurements in QFT.

\section{The detector model}\label{sec:detector}

In this work we will consider a  version of the  Unruh-DeWitt detector model that is suitable for curved backgrounds. In QFT in curved spacetimes, one is often forced to restrict the analysis to globally hyperbolic manifolds, which are equipped with a Lorentzian metric $g_{\mu\nu}$ \cite{wald2010general}. It can be shown, at least in some contexts (e.g. locally covariant AQFT \cite{haag2012local}) that one can formulate a quantum theory of fields propagating in this type of manifolds and the fields will generally obey some hyperbolic equation of motion \cite{frankwave}. In the case of a scalar field one possible such equation is the minimally coupled Klein-Gordon equation
\begin{align}
    \nabla_\mu\nabla^\mu \hat\phi(\mathsf{x})-m^2\hat\phi(\mathsf{x})=0.
\end{align}

To define a detector model in general relativistic set-ups, it is convenient to introduce the interaction Hamiltonian defining the coupling between detector and field in terms of Hamiltonian densities, which are covariant objects \cite{PhysRevD.101.045017}.

Now, consider a  quantum system coupling to the field through the following Hamiltonian density in the interaction picture:
\begin{align}
\hat{\mathfrak{h}}(\mathsf x)=\sqrt{|g|}\hat h (\mathsf x),
\end{align}
where the \textit{Hamiltonian weight} $\hat h (\mathsf x)$ is defined as
\begin{align}\label{h4}
    \hat h (\mathsf x)\coloneqq\lambda \Lambda(\mathsf x) \hat J(\mathsf x)\otimes\hat \phi(\mathsf x).
\end{align}
 $\Lambda$ is a space-time function of compact support, $\hat{J}$ is a current operator associated with the detector and $\lambda$ is a coupling constant. This Hamiltonian density generalizes many particle detector models in the literature. 

From this Hamiltonian density one can define a time dependent Hamiltonian for the joint system as
\begin{align}\label{h5}
    \hat H(t)=\int_{\mathcal{E}(t)}\!\!\!\!\!\d\mathcal{E} \; \hat h(\mathsf x),
\end{align}
where $\mathcal{E}(t)$ is a one-parameter family of spacelike surfaces.
The parameter $t$ is given by the values of a global function, $\mathscr{T}(\mathsf{x})$, whose level curves can be taken to be the planes of simultaneity  of some observer. Under some assumptions \cite{PhysRevD.101.045017,2020broken}, $t$ can be chosen to be the actual proper time of the detector.  Finally, $\d\mathcal{E}$ is a shorthand for the family of induced measures such that
\begin{align}
   \int_{\mathcal{E}(t)}\!\!\!\!\! \d\mathcal{E} :=\int \d V \delta(\mathscr{T}(\mathsf{x})-t)
\end{align} in the surfaces $\mathcal{E}(t)$, where $\d V=\sqrt{|g|} \d x^{n}$, $n$ being the dimension of the manifold. 

 Note that the operator $\hat J$ can typically depend on a privileged foliation associated with the proper time of the center of mass of the detector, which we will denote $\tau(\mathsf{x})$. In fact, in the usual UDW prescription, 
\begin{align}
    \hat J(\mathsf{x})=\hat{\mu}(\tau(\mathsf{x})),
\end{align}
where $\hat\mu$ is the monopole operator given by $\hat\mu(\tau)=e^{\ii\Omega \tau}\hat\sigma^++e^{-\ii\Omega \tau}\hat\sigma^-$.
The family of models \eqref{h5} with the standard UDW prescription for the operators acting over the detector reduces to the standard UDW model when the time evolution is prescribed by an observer fiduciary to the detector, that is $\mathscr{T}(\mathsf{x})=\tau(\mathsf{x})$. In that case 
\begin{align}
    &\nn\hat H(t)=\lambda\int_{\mathcal{E}(t)}\!\!\!\!\!\d\mathcal{E} \;  \Lambda(\mathsf x) \hat{\mu}(\tau(\mathsf{x}))\otimes\hat \phi(\mathsf x)\\
    &=\lambda\hat{\mu}(t)\otimes\int_{\mathcal{E}(t)}\!\!\!\!\!\d\mathcal{E} \;  \Lambda(\mathsf x) \hat \phi(\mathsf x).
\end{align}
 One can regard the current $\hat J$ as a property of the detector and the function $\Lambda$ as a property of the detector-field interaction, introduced phenomenologically.  This type of detector models in curved space-times have been introduced very recently with finite dimensional detectors \cite{PhysRevD.101.045017}. Hamiltonian \eqref{h5}  accounts for general detectors and general scalar fields in curved spacetimes, and so we will use it as a starting point to analyse general features that concern all types of detector models.

\section{ faster-than-light signaling in detector models}\label{sec:faster}
This section is devoted to provide a general argument concerning the existence of faster-than-light signalling in measurement schemes  modeled with particle detector models. 

In the underlying quantum field theory superluminal signalling is prevented through the microcausality axiom, which states that the field operators commute in spacelike separation, i.e.
\begin{align}
    [\hat{\phi}(\mathsf{x}),\hat{\phi}(\mathsf y)] =0,
\end{align}
if $\mathsf{x}$ and $\mathsf{y}$ are spacelike separated.  The link between the microcausality axiom and signalling stems from the notion of statistical independence of measurements. Consider for example two commuting observables $[\hat{A},\hat{B}]=0$ of a closed quantum system. A non-selective measurement of observable $\hat{A}$ does not affect the expectation value of $\hat{B}$. Indeed, let $\hat{\rho}_0$ be the initial state. Consider the spectral decomposition of $\hat{A}$, i.e., $\hat{A}=\sum_a a \hat{P}_a$. After a non-selective measurement of $\hat A$ the state of the system, denoted as $\hat{\rho}|_\textsc{a}$, is given by
\begin{align}
    \hat{\rho}|_\textsc{a}= \sum_a \hat{P}_a \hat{\rho}_0 \hat{P}_a \label{nonsel}
\end{align}
Since $[\hat{A},\hat{B}]=0$, it holds that $[\hat{P}_a,\hat{B}]=0 \,\,\,\forall a$. Making use of this, we can easily see that 
\begin{align}
    \text{tr}(\hat{B}\hat{\rho}|_\textsc{a})=\text{tr}(\hat{B}\hat{\rho}_0) \label{no}
\end{align}
which means that the expectation value of $\hat{B}$ does not depend on a \textit{non-selective} measurement (i.e., a measurement of which the outcome is not known) having happened. In the context of quantum field theory, where the commuting $\hat{A}$ and $\hat{B}$ are explicitly associated with spacelike separated regions, microcausality guarantees condition \eqref{no}, which implies no-signaling in spacelike separation.  

It is important to clarify that in the case of selective measurements microcausality does not guarantee statistical independence, i.e., the statistics of $\hat{B}$ will generally depend on the outcome\footnote{Above we denoted as $\hat{\rho}|_\textsc{a}$ the state of the system given that the observable A has been measured but the value of the outcome A is not known, i.e., the mixture \eqref{nonsel}, which corresponds to a non-selective measurement. On the other hand, in a selective measurement a particular outcome A is `selected' and the state of the system given this outcome is simply $\hat{\rho}|_a=\ket{a}\bra{a}$. In this case, one can see that the analogue of \eqref{no} does not hold, i.e.,  $\text{tr}(\hat{B}\hat{\rho}|_a)\neq\text{tr}(\hat{B}\hat{\rho}_0)$.} A . It is well known that quantum field theory permits outcome-outcome correlations in spacelike separation \cite{redhead1995more}. It is also well-understood that outcome-outcome correlations do not lead to superluminal signaling. Rather, they are a consequence of the fact that the state of the field (for example, a thermal state, or  the vacuum) can display entanglement and classical correlations even between spacelike separated regions~\cite{summers1987}.  

In the context of detector models, measurements are to be implemented through the dynamical coupling of a detector to the quantum field. The dynamical coupling is implemented through a unitary operator that acts over joint detector-field system, which is meant to describe the evolution of the joint system from an ``in'' spacetime region to an ``out'' spacetime region. The detector system is supposed to be initially uncorrelated with the field in the ``in region''. After the interaction has finished, the statistics of the detector system are analyzed.  Effectively, this implies to analyze some subset of the statistics of the field since the detector and the field are now correlated due to the interaction.

In this section we will make use of the most general possible (linear) detector model of the family of models described by the Hamiltonian density~\eqref{h4}. Further, if we are to restrict our set of measurements of the field to this kind of detector-based protocols, it is necessary to define signalling  in terms of interactions between several detectors. 

Let us study then the dynamics of a set of independent detectors interacting with the same quantum field.
First, consider two detectors that couple to the same quantum field undergoing an interaction generated by the Hamiltonian density
\begin{align}
    &\nn\hat h(\mathsf x)=\hat h_\textsc{a}(\mathsf x)+\hat h_\textsc{b}(\mathsf x)\\
    &\nn=\lambda_\textsc{a} \Lambda_\textsc{a}(\mathsf x) \hat J_\textsc{a}(\mathsf x)\otimes\openone_{\textsc{b}}\otimes\hat \phi(\mathsf x)\\
    &+\lambda_\textsc{b} \Lambda_\textsc{b}(\mathsf x) \openone_{\textsc{a}}\otimes\hat J_\textsc{b}(\mathsf x)\otimes\hat\phi(\mathsf x).
\end{align}

This Hamiltonian density generates a joint Hamiltonian for the joint system of the form 
\begin{align}
    \hat H({t})=\hat H_\textsc{a}({t})+\hat H_\textsc{b}({t})
\end{align}
where 
\begin{align}\label{hdos}
    \hat H_{\textsc{a},\textsc{b}}({t})=\int_{\mathcal{E}({t})}\!\!\!\!\!\d\mathcal{E}\;   \hat h_{\textsc{a},\textsc{b}}(\mathsf x).
\end{align}
 Note that this Hamiltonian generates evolution with respect to the same parameter ${t}$ for both detectors. Although we will not concern ourselves with this in the present work, since it has already been studied in \cite{2020broken}, it is clear that one needs to properly reparametrize the local Hamiltonians to generate time translations with respect to the same parameter, which in general cannot correspond to the proper time of both detectors.

In order to analyze causal relations between detectors, we need first to define causal relations between subsets of spacetime.  We are going to define the causal order relations between two detectors  A and B in terms of the causal order relations between the spacetime regions in which their interaction is supported: $\mathcal{O}_{\textsc{a}}=\text{supp}(\Lambda_{\textsc{a}})$, \mbox{$\mathcal{O}_{\textsc{b}}=\text{supp}(\Lambda_{\textsc{b}})$}.   Here we will consider only spacetime smearing functions $\Lambda$ that are compactly supported and so they single out compact regions $\mathcal{O}=\text{supp}(\Lambda)$ for the detector-field interaction \eqref{h4}.

Given a globally hyperbolic spacetime, the future lightcone of a compact region $\mathcal{O}$, $\mathcal{J}^{+}(\mathcal{O})$, is the set of all points that lay in the causal future of some point of $\mathcal{O}$. Similarly, $\mathcal{J}^{-}(\mathcal{O})$, the causal past of a region $\mathcal{O}$, is the set of all points that lay in the causal past of $\mathcal{O}$.

\begin{itemize}
\item We say that A and B are causally orderable if $\mathcal{J}^-(\mathcal{O}_\textsc{a})\cap\mathcal{O}_\textsc{b}$ or  $\mathcal{J}^-(\mathcal{O}_\textsc{b})\cap\mathcal{O}_\textsc{a}$ are empty.  
   
    \item  We say that A and B are spacelike separated if
         $(\mathcal{J}^+(\mathcal{O}_\textsc{a})\cup\mathcal{J}^-(\mathcal{O}_\textsc{a}))\cap\mathcal{O}_\textsc{b}$ or $(\mathcal{J}^+(\mathcal{O}_\textsc{b})\cup\mathcal{J}^-(\mathcal{O}_\textsc{b}))\cap\mathcal{O}_\textsc{a}$ are empty. Notice that this is a particular case of causally orderable. 
    \item   Finally, we have that if $\mathcal{O}_\textsc{b}\subset\mathcal{J}^+(\mathcal{O}_\textsc{a})\setminus \mathcal{O}_\textsc{a}$, and \mbox{$\mathcal{O}_\textsc{a}\subset\mathcal{J}^-(\mathcal{O}_\textsc{b})\setminus \mathcal{O}_\textsc{b}$}, we say that A causally precedes B. Notice that this is a particular case of causally orderable since although $\mathcal{J}^-(\mathcal{O}_\textsc{b})\cap\mathcal{O}_\textsc{a}=\mathcal{O}_\textsc{a}\neq \emptyset$, it holds that $\mathcal{J}^-(\mathcal{O}_\textsc{a})\cap\mathcal{O}_\textsc{b}=\emptyset$.
\end{itemize}

\begin{figure}
    \includegraphics[width=0.5\textwidth]{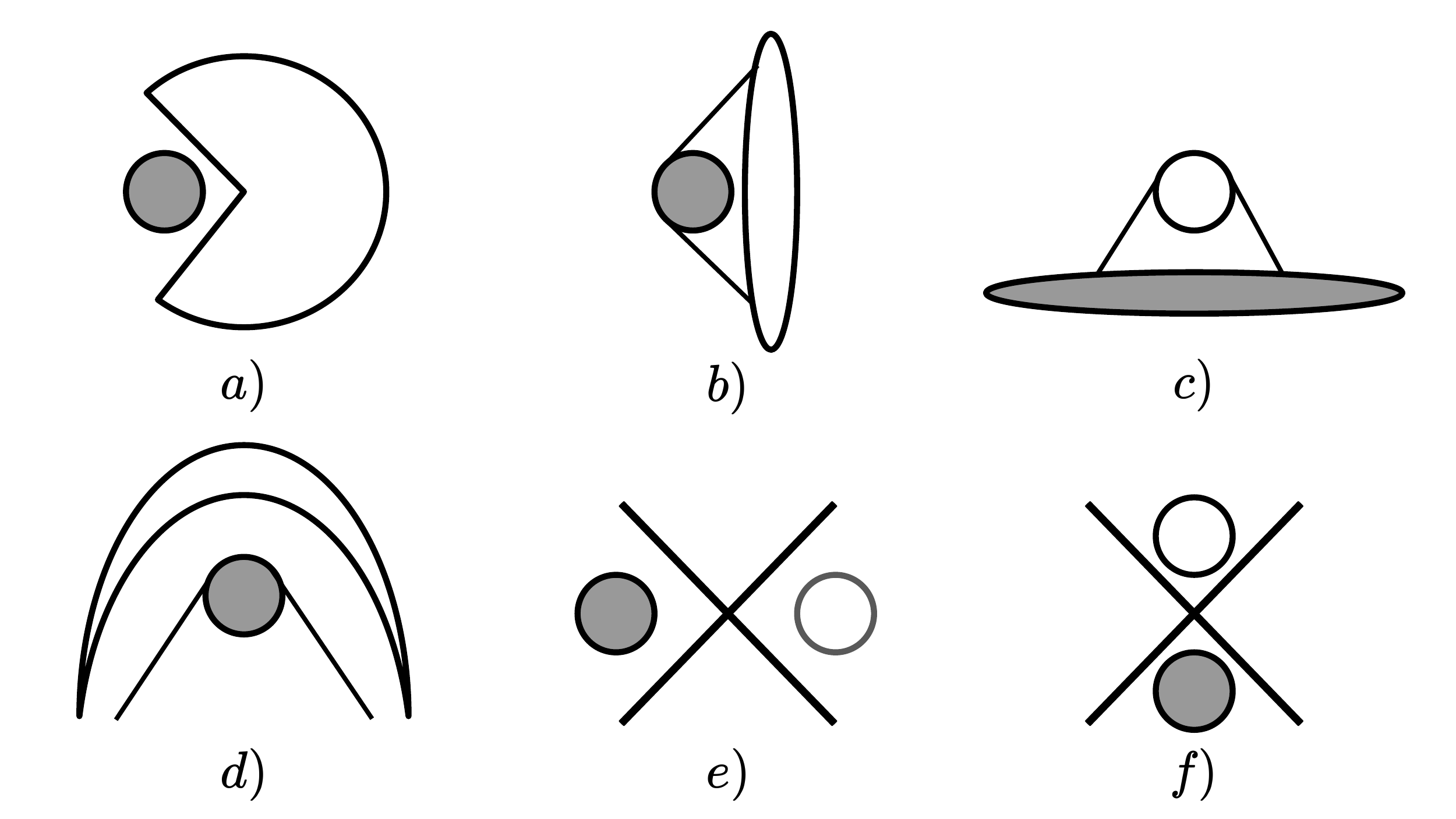}
    \caption{Causal relations between simply connected, non intersecting sets (grey and white) in two spacetime dimensions. Black lines represent the future or past lightcone of the sets or points between them.
    $a),b)$: Examples of non causally orderable sets. 
    $c),d)$: Examples of sets that are causally orderable, but that do not causally precede each other according to our definition. 
    $e)$: Spacelike separated sets.
    $f)$: Example of a set causally preceding another set.}
    \label{fig:figuritas}
\end{figure}
  Specifically, we have defined $\mathcal{O}_\textsc{a}$ to precede $\mathcal{O}_\textsc{b}$ if for every observer all the events in $\mathcal{O}_\textsc{a}$ precede any event in $\mathcal{O}_\textsc{b}$, that is, $\mathcal{O}_\textsc{a}$ ``comes first" for all observers.

    These are covariant statements that are independent of the observer, but one can also define causal relations with respect to a particular foliation $\mathscr{T}(\mathsf{x})$. We say that A precedes B with respect to $\mathscr{T}(\mathsf{x})$ if $\mathscr{T}(\mathsf{x})<\mathscr{T}(\mathsf{y})$ for all $\mathsf{x}\in\mathcal{O}_\textsc{a}$ and for all $\mathsf{y}\in\mathcal{O}_\textsc{b}$.  The two notions are linked by the following facts:
     \begin{itemize}   
    \item  If A and B are causally orderable (since  $\mathcal{O}_\textsc{a}$ and $\mathcal{O}_\textsc{b}$ are  compact) one precedes the other with respect to some foliation.
    \item   If $\mathcal{O}_\textsc{a}$ and $\mathcal{O}_\textsc{b}$  are spacelike separated (since they are compact)  then there are at least two foliations such that A precedes B with respect to one and such that B precedes A with respect to the other.
        \item If A causally precedes B, then A  precedes B with respect to all foliations.
    \end{itemize}

    Given these definitions of causal relations, we can analyze further the implications of the microcausality axiom in detector physics. The Hamiltonians defined by \eqref{hdos} are defined respect to some time function $\mathscr{T}(\mathsf{x})$, so the two detectors will naturally have causal relations with respect to the foliation defined by its level curves. If the underlying field theory were not relativistic, we would expect that different foliations give rise to different dynamics for spacelike separated detectors, because in that case the order in which the measurements  are done would typically matter. This is exactly what is to be avoided in a relativistic theory, and in the following we will examine this condition in detector models departing from the microcausality condition of the underlying QFT. 
    
Now, recall that the microcausality axiom in curved spacetimes implies that, for two compactly supported spacetime functions $m(\mathsf{x})$ and $l(\mathsf{x})$, 
\begin{align}
    \int\d\mathsf{V}\int\d\mathsf{V}'\,
     l(\mathsf{x})m(\mathsf{y})[\hat{\phi}(\mathsf{x}),\hat{\phi}(\mathsf y)]=0
\end{align}
where $\d\mathsf{V}=\d\mathsf{x}^{n}\sqrt{|g|}$ and $\d\mathsf{V}'=\d\mathsf{y}^{n}\sqrt{|g'|}$, if the supports of $l$ and $m$ are spacelike separated.
Therefore, the microcausality axiom implies that 
\begin{align}
     \big[\hat h_{\textsc{a}}(\mathsf x),\hat h_{\textsc{b}}(\mathsf y)\big]=0
\end{align}
if $\Lambda_{\textsc{a}}$ and $\Lambda_{\textsc{b}}$ have spacelike separated supports. This in turn implies that
\begin{align}
     \big[\hat H_{\textsc{a}}({t}),\hat H_{\textsc{b}}({t}')\big]=0
\end{align}
for all ${t},{t}'$. 

 The joint evolution in the of the detectors and the field can be described as a unitary operator acting over the joint state of the system. That is, if $\hat\rho_{\text{initial}}$ is the density operator describing the state of the field-detectors system before the interactions are switched-on (respect to the parameter ${t}$). The notation A$+$B indicates that the operator accounts for the interaction of the two detectors, whereas $\hat S_{\textsc{a},\textsc{b}}$ will denote the scattering matrices associated with the individual interactions generated by the individual interaction Hamiltonians. The total state in the asymptotic future will be given by the transformation
\begin{align}
    \hat\rho_{\text{final}}=\hat S_{\textsc{a}+\textsc{b}}\;\hat\rho_{\text{initial}}\;{\hat S}^{\dagger}_{\textsc{a}+\textsc{b}}
\end{align}
where $\hat S_{\textsc{a}+\textsc{b}}$ is the so-called scattering operator\footnote{It is common in the UDW literature to denote the evolution operator by $\hat{U}$. In this work, however, we prefer to denote it with $\hat{S}$ to emphasize the fact that these maps represent scattering operators and we adopt a notation analog to, e.g., \cite{bostelmann2020impossible} .}. The scattering operator is unitary and can be formally written as the Dyson series
\begin{align}
  &\nn\hat S_{\textsc{a}+\textsc{b}} =\sum_{n}\left(\frac{-\ii}{\hbar}\right)^n\frac{1}{n!} \int_{-\infty}^{\infty}\dots\int_{-\infty}^{\infty} \d {t}^{n}\\
&  \times \mathcal{T}(\hat H_{\textsc{a}}({t}_1)+\hat H_{\textsc{b}}({t}_1)\dots\hat H_{\textsc{a}}({t}_n)+\hat H_{\textsc{b}}({t}_n)).
\end{align}

Intuitively, we would like to ensure that if two detectors A and B are coupled to the field in spacelike separation, one cannot conclude whether the other one is coupled to the field or not.
Therefore, a minimum non-signaling requirement would be that if A interacts first with the quantum field in any foliation, i.e. if B is not in the causal future of A, then all expectation values of observables of detector B should not depend on magnitudes of detector A, e.g. the coupling constant $\lambda_\textsc{a}$. If the expectation values of observables of B  depend on $\lambda_\textsc{a}$, then its value could be used to encode, and then signal information.

It is not a priori obvious why the causal behaviour of the underlying QFT, e.g. the microcausality axiom, would guarantee the causal behaviour of detectors. However, we will see that this is guaranteed under some conditions~\cite{PhysRevD.92.104019}.  As we will show below, in the context of particle detector models, faster-than-light signalling is prevented if the joint scattering matrix factorizes when the detectors are causally orderable. In particular, if B does not intersect with the past of A, we would have
\begin{align}\label{caufac}
    \hat S_{\textsc{a}+\textsc{b}}=\hat S_{\textsc{b}}\hat S_{\textsc{a}}.
\end{align}
 We will refer to this property as causal factorization.

To see that causal factorization prevents acausal signalling, consider the local statistics of the detector A, given by the partial trace
\begin{align}
    \hat\rho_{\textsc{a}}=\tr_{\textsc{b},\phi}(\hat S_{\textsc{a}+\textsc{b}}\;\hat\rho_{\text{initial}}\;{\hat S}^{\dagger}_{\textsc{a}+\textsc{b}}).
\end{align}
Now, if causal factorization holds, then 
\begin{align}
    \hat\rho_{\textsc{a}}=\tr_{\textsc{b},\phi}(\hat S_{\textsc{b}}\hat S_{\textsc{a}}\;\hat\rho_{\text{initial}}\;{\hat S}^{\dagger}_{\textsc{a}}{\hat S}^{\dagger}_{\textsc{b}}).
\end{align}
But $\hat S_{\textsc{b}}$ depends only on operators acting over the subspaces associated with the field and the detector B, therefore it can be permuted within the partial trace:
\begin{align}
    &\nn\hat\rho_{\textsc{a}}=\tr_{\textsc{b},\phi}(\hat S_{\textsc{a}}\;\hat\rho_{\text{initial}}\;{\hat S}^{\dagger}_{\textsc{a}}{\hat S}^{\dagger}_{\textsc{b}}\hat S_{\textsc{b}})\\
   &=\tr_{\textsc{b},\phi}(\hat S_{\textsc{a}}\;\hat\rho_{\text{initial}}\;{\hat S}^{\dagger}_{\textsc{a}}) .
\end{align}
Therefore, we have shown that if causal factorization holds, there is no local (space-time compact) measurement carried though a detector interaction that can be used to receive signals from another detector outside the causal past of such interaction. 

Note that in the particular case where A and B are spacelike separated, then causal factorization implies
\begin{align}
    \hat S_{\textsc{a}+\textsc{b}}=\hat S_{\textsc{b}}\hat S_{\textsc{a}}=\hat{S}_{\textsc{a}}\hat S_{\textsc{b}},
\end{align}
which implies that neither detector A can signal to detector B nor detector B can signal to detector A, that is, it prevents faster than light signaling.

We provide a proof of causal factorization in appendix \ref{app:factorization}, which relies heavily on the microcausality condition fulfilled by the field operators. It is rather intuitive why condition \eqref{caufac} should hold  if, e.g. A precedes B respect to the concrete foliation in which the interaction has been defined, as the unitary evolution factorizes by construction. This can be used to argue that the factorization will be independent of the foliation if A  causally precedes B in the sense given at the beginning of this section. The proof is also simple if A and B are spacelike separated, in which case the factorization also holds independently of the foliation. What is less trivial, however, is that the factorization holds if the detectors are causally orderable, which is a covariant statement that does not depend on the foliation either.

In conclusion, causal factorization prevents faster-than-light signalling, as far as only two detectors are involved. 
The result can be extended to some limited scenarios with many detectors. For example, if one has more than two detectors, say $\text{A},\text{B}_1,\dots,\text{B}_N$, one can always define the collection of all the detectors that are not A as a single detector $\text{A}^c$. If all the detectors in $\text{A}^c$ and A are causally orderable, with A preceding the rest, then again causal factorization will hold and
\begin{align}
\hat S_{\Sigma \textsc{b}_i+ \textsc{a}}=\hat S_{\Sigma \textsc{b}_i}\hat S_{\textsc{a}}=\hat S_{\textsc{a}^c}\hat S_{\textsc{a}}
\end{align}
and the measurements on A will not be affected by the other detectors.

It could be tempting to claim that this implies that the signals sent by a detector can only reach other detectors in the causal future of its interaction region. Indeed, causal factorization ensures this as long as we consider schemes involving two detectors. Obviously, if a detector B can only receive signals from its causal past, then another single detector A can only send signals to B if B is in the causal future of A. 

However, if more than two detectors are involved, then causal factorization does not solve all the possible frictions that the detector models can have with relativistic causality. We deal with this in the next section.

\section{ ``Impossible measurements'' and Superluminal propagation of initial data}\label{sec:propagation}

We have defined signalling so far as the transmission of information between detectors through their interaction with the field. We have seen that, in a two-detector scenario, a detector localized in some region is irrelevant for 
another detector localized in its causal complement, which means that the detector only influences, in some sense, its own causal future.

However, as pictured in Sorkin's impossible measurements paper~\cite{sorkin1993impossible},  there are subtleties associated with the detectors not being in a definite causal ordering when considering more than two measurements. Namely, even if the response of the detector A cannot be influenced by the detector B, the influence of detector A over B can still carry information about events that happened outside the causal past of B, which is obviously not acceptable.

In order to understand how Sorkin's problem can manifest in measurement's models with particle detectors, we shall first analyze a different kind of signalling in which the information is not encoded in the interaction, but in the initial state of the system.

Indeed, a detector can also be thought of as a repeater, that is, given some initial state of the field (possibly coming from another interaction), the detector can register the initial data and propagate it back to the field. In this case, one may fear that a detector can re-emit information in a non-causal manner. In this subsection we will prove that this is indeed a reasonable concern, since superluminal propagation of initial data is a widespread phenomenon when considering non-relativistic systems.  

\begin{figure}
\centering
\includegraphics[width=0.45\textwidth]{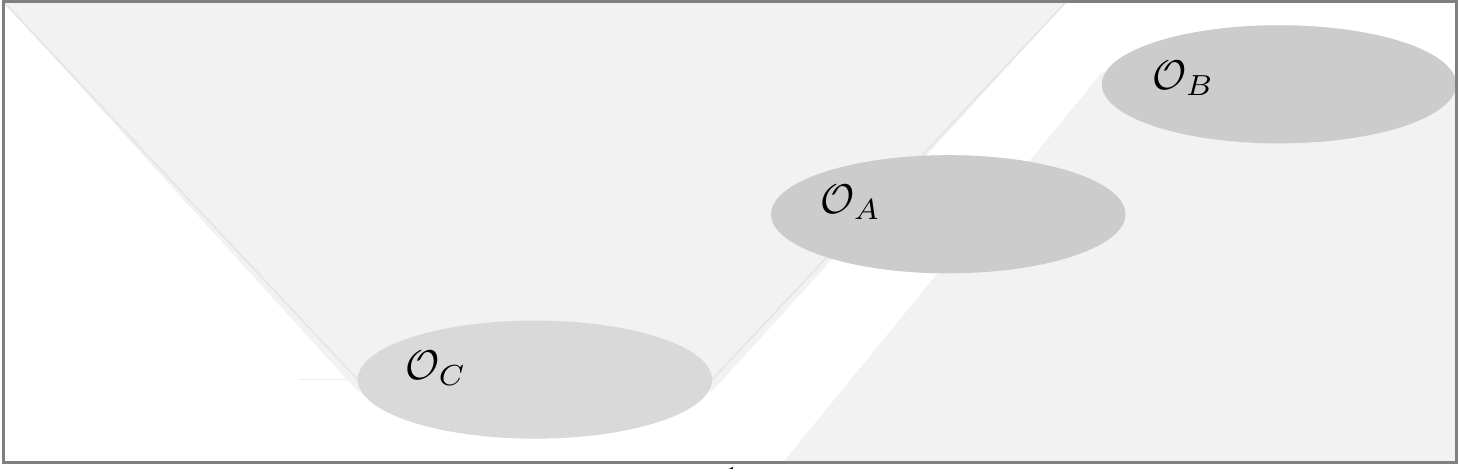}
\caption{ \hspace*{-0.2cm} Two detectors A and B are coupled to the field over regions $\mathcal{O}_{\textsc{a},\textsc{b}}$. The initial data are encoded in the field state through a unitary intervention over region $\mathcal{O}_{\textsc{c}}$. Notice that region $\mathcal{O}_\textsc{a}$ is partially invading the past and future lightcones of regions $\mathcal{O}_\textsc{b}$ and $\mathcal{O}_\textsc{c}$ respectively.}
\label{sorkin2}
\end{figure}

 For instance, one could imagine a scenario in which a detector A partially precedes and is partially spacelike separated from a second detector B (see figure \ref{sorkin2}). Consider that the state of the system is initially given by
\begin{align}
    \hat\rho_{\text{initial}}=\hat\rho_\textsc{a}\otimes\hat\rho_\textsc{b}\otimes e^{\ii\lambda_f \hat\phi(f)}\hat\rho_\phi e^{-\ii\lambda_f \hat\phi(f)}
\end{align}
where $\hat\rho_{\textsc{a},\textsc{b},\phi}$ are arbitrary states of the detector A, B and the quantum field respectively, and $\hat\phi(f)$ is a smeared field operator which is compactly supported in region\footnote{One can think of this as a third party Charles encoding information in the field in region through a spacetime localized unitary action in region $\mathcal{O}_\textsc{c}$. } $\mathcal{O}_\textsc{c}$, spacelike separated from B, but not from A. If $\mathcal{O}_\textsc{c}$ is spaceilike separated from B, the local statistics of B should not be affected by the value of the constant $\lambda_f$, otherwise detector A would be acting as an agent for superluminal signalling.

More generally, one can consider the case in which the initial state of the detectors plus field has the form
\begin{align}\label{unita}
    \hat\rho_{\text{initial}}=\hat U \hat\rho_{0}\hat U^{\dagger},
\end{align}
where $\hat\rho_{0}$ is an arbitrary reference state of the joint system, and $\hat U =\openone_\textsc{a}\otimes\openone_\textsc{b}\otimes\hat U_{\phi}$ is an arbitrary unitary acting on the field's Hilbert space, so that $\hat U_{\phi}$ is localized in $\mathcal{O}_\textsc{c}$ (contained in the causal complement of the interaction region $\mathcal{O}_\textsc{b}$). It is clear that
\begin{align}
    [\hat U,\hat S_\textsc{b}]=0.
\end{align}
$\hat U$ can be thought of as encoding a set of initial data\footnote{We can always thing without loss of generality that the action of $\hat U$ is localized in a subset of a Cauchy surface in the causal past of  $\mathcal{O}_\textsc{c}$}. The statistics of detector B can only depend on $\hat U$ if detector A's interaction region overlaps with the causal past of B and $\hat{S}_\textsc{a}$ does not commute with $\hat U$ (e.g., as shown in Fig.~\ref{sorkin2}).  To avoid superluminal signalling, it should hold that the local statistics of B do not depend on the choice of $\hat U$, i.e.
\begin{align}\label{firstcon}
    \hat\rho_{\textsc{b}}=\tr_{\textsc{a},\phi}(\hat S_{\textsc{a}+\textsc{b}}\hat U\;\hat\rho_{0}\;{\hat U}^{\dagger}{\hat S}^{\dagger}_{\textsc{a}+\textsc{b}})=\tr_{\textsc{a},\phi}(\hat S_{\textsc{a}+\textsc{b}}\;\hat\rho_{0}\;{\hat S}^{\dagger}_{\textsc{a}+\textsc{b}}).
\end{align}

 Further, since B is localized (at least partially) in the future of A, and it is spacelike separated from the set of initial data implemented by $\hat U$, B cannot be fully contained in the causal past of A. We conclude that A does not causally precede B, in the terminology of the last section. 

Imposing condition \eqref{firstcon} for all initial density operators is equivalent to
\begin{align}
    \tr_{\textsc{a},\phi}(\hat V\;\hat\sigma\;\hat{V}^\dagger)=\tr_{\textsc{a},\phi}(\hat\sigma)
\end{align}
for any arbitrary density operator $\hat \sigma$,
where $\hat V$ is a unitary given by
\begin{align}
    \hat V={\hat S}_{\textsc{a}+\textsc{b}}\hat U{\hat S}^{\dagger}_{\textsc{a}+\textsc{b}}.
\end{align}
 This implies that if $\hat D_\textsc{b}$ is an operator acting on detector B (i.e. it commutes with the field operators and with the operators acting on detector A) then
 \begin{align}
     \tr(\hat{V}^\dagger\hat D_\textsc{b}\hat{V}\hat \sigma)=\tr(\hat D_\textsc{b}\hat \sigma)
 \end{align}
 for all $\hat \sigma$. For our purposes, this implies that
 \begin{align}
     \hat{V}^\dagger\hat D_\textsc{b}\hat{V}=\hat D_\textsc{b},
 \end{align}
 or equivalently
  \begin{align}\label{seccond}
     [\hat D_\textsc{b},\hat{V}]=0
 \end{align}
 for all operators acting over detector B. Assuming that A precedes B, the connection with the propagation of initial data is more clear when one uses causal factorization. Then, ${\hat S}_{\textsc{a}+\textsc{b}}={\hat S}_{\textsc{b}}{\hat S}_{\textsc{a}}$ and condition \eqref{seccond} can be written as
  \begin{align}\label{thirdcond}
     [{\hat S}^{\dagger}_{\textsc{b}}\hat D_\textsc{b}{\hat S}_{\textsc{b}},{\hat S}_{\textsc{a}}\hat{U}{\hat S}^{\dagger}_{\textsc{a}}]=0,
 \end{align}
 for all unitaries in the causal complement of B. If we think of ${\hat S}^{\dagger}_{\textsc{b}}\hat D_\textsc{b}{\hat S}_{\textsc{b}}$ as an induced operator acting on the field localized in region B and of  ${\hat S}_{\textsc{a}}\hat{U}{\hat S}^{\dagger}_{\textsc{a}}$ as the evolution of the initial data given by interaction A, we can interpret condition \eqref{seccond} as that the interaction A does not propagate initial data superluminally, since the propagated data still lays within the causal complement of region B. This condition is related to the unitary restriction of the condition  discussed in \cite{borsten2019impossible}, but more general in the sense that allows for auxiliary degrees of freedom representing the devices used to implement the measurement.

 The relevant question now is whether condition \eqref{seccond} holds for general detector models. Unfortunately the answer is generally negative. It is easy to corroborate using perturbation theory that the localization region of ${\hat S}_{\textsc{a}}\hat{U}{\hat S}^{\dagger}_{\textsc{a}}$ is not the causal future of $\hat{U}$, but the causal future of A. Indeed, using Dyson's expansion
\begin{align}
   &\nn\hat S_{\textsc{a}}\;\hat U\;{\hat S}^{\dagger}_{\textsc{a}}=\hat U -\frac{\ii}{\hbar}\int\d V 
[\hat h_{\textsc{a}}(\mathsf{x}),\hat U]\\
& \nn -\frac{1}{2\hbar^2}\int\d V \int\d V' \mathcal{T} \left[\hat h_{\textsc{a}}(\mathsf{x}), [\hat h_{\textsc{a}}(\mathsf{y}),\hat U]\right]\\
& +\mathcal{O}(\lambda_\textsc{a}^3).
\end{align}
 If we pay attention to the first term, which is given by the density
 \begin{align}
     [\hat h_{\textsc{a}}(\mathsf{x}),\hat U]=\lambda_\textsc{a} \Lambda_\textsc{a}(\mathsf x)\hat J_\textsc{a}(\mathsf x)\otimes \openone_{\textsc{b}}\otimes[\hat\phi(\mathsf x),\hat U],
 \end{align}
 we realize that microcausality ensures that no $\mathsf{x}$ outside the lightcone of $\hat U$ can contribute to the integral.  This means that regardless of the localization of region B, the leading order propagation of initial data is still localized in the lightcone of $\hat U$ and the propagation is causal.
 
 Now, at second order, the contribution will be given by the kernel
 \begin{align}
     \left[\hat h_{\textsc{a}}(\mathsf{x}), [\hat h_{\textsc{a}}(\mathsf{y}),\hat U]\right]
 \end{align}
 where the time-ordering is implemented considering that  $\mathsf{y}$ precedes $\mathsf{x}$ respect to the foliation $\mathscr{T}(\mathsf{x})$. Because of microcausality, $\mathsf{y}$ will also be constrained to lie within the lightcone of $\mathcal{O}_\textsc{c}$, but $\mathsf{x}$ can be anywhere. One can use Jacobi's identity to expand this kernel as follows
  \begin{align}
     &\nn\left[\hat h_{\textsc{a}}(\mathsf{x}), [\hat h_{\textsc{a}}(\mathsf{y}),\hat U]\right]\\
     &= \left[[\hat h_{\textsc{a}}(\mathsf{x}),\hat h_{\textsc{a}}(\mathsf{y})],\hat U\right]+ \left[\hat h_{\textsc{a}}(\mathsf{y}), [\hat h_{\textsc{a}}(\mathsf{x}),\hat U]\right],
 \end{align}
 such that   $\mathsf{x}$ has to lie in the lightcone of the initial data for the second term not to vanish, but the first one will not generally vanish when $\mathsf{x}$ is outside the lightcone of $\mathcal{O}_\textsc{c}$.
 
 One can see that in general, unless $[\hat h_{\textsc{a}}(\mathsf{x}),\hat h_{\textsc{a}}(\mathsf{y})]=0$ when $\mathsf{x}$ and $\mathsf{y}$ are spacelike separated, the propagation will not be causal anymore. Similar results were found  in \cite{2020broken} when addressing violations of relativistic covariance.
 
 Indeed, one can further expand the commutator of the Hamiltonian densities as 
 \begin{align}\label{denscomm}
     &\nn[\hat h_{\textsc{a}}(\mathsf{x}),\hat h_{\textsc{a}}(\mathsf{y})]\\
    &\nn =\lambda_\textsc{a}^2 \Lambda_\textsc{a}(\mathsf x) \Lambda_\textsc{a}(\mathsf y)[\hat J_\textsc{a}(\mathsf x),\hat J_\textsc{a}(\mathsf y)]\otimes\openone_{\textsc{b}}\otimes\hat\phi(\mathsf x)\hat\phi(\mathsf y)\\
    &+\lambda_\textsc{a}^2 \Lambda_\textsc{a}(\mathsf x) \Lambda_\textsc{a}(\mathsf y)\hat J_\textsc{a}(\mathsf x)\hat J_\textsc{a}(\mathsf y)\otimes\openone_{\textsc{b}}\otimes[\hat\phi(\mathsf x),\hat\phi(\mathsf y)].
 \end{align}
 Again, microcausality ensures that the second term in \eqref{denscomm} vanishes in spacelike separation, but the first one will not vanish, nor will commute with $\hat{U}$ in general , unless $[\hat J_\textsc{a}(\mathsf x),\hat J_\textsc{a}(\mathsf y)]=0$ in spacelike separation. In general it is not difficult to argue (following a similar combinatoric procedure as in~\cite{2020broken}, together with a recursive use of Jacobi's identity) that if the interaction Hamiltoinan density of A is microcausal (for example for a pointlike detector), the propagation of initial data is causal in all orders in perturbation theory.
 
 If this condition holds, it means that either all points in $\text{supp}\, \Lambda_\textsc{a}$ are causally connected (which is only possible for a pointlike detector) or that the detector is a relativistic field. Since by assumption the system is non-relativistic and generally smeared, we conclude that the detector's dynamics carry superluminal propagation of initial data at second order in perturbation theory.
 
 Note that since  for point-like detectors there is not superluminal propagation, one can disregard this kind of faster-than-light signalling for ``small enough'' detectors. Whether a detector is small or not will depend, of course, of the parameters of the problem.

 The preceding discussion provides a dynamical interpretation of the impossible measurements problem, in the sense that it links superluminal signalling with superluminal propagation within the device that is implementing the measurement. 
 It is clear then, that if the detector is a relativistic quantum field then there is not superluminal propagation of initial data under some assumptions in the dynamics of the coupling, as it is shown in full rigor in \cite{bostelmann2020impossible}. In our case, however, we have to understand this kind of faster-than-light signalling as a fundamental feature of non-relativistic particle detector models that restricts their usage to regimes where these superluminal features are negligible or irrelevant for the results at hand. 
 
 \section{Impossible measurements with weakly coupled detectors}\label{sec:perturbation}
 
 We have seen that faster-than-light signalling is present in smeared non-relativistic particle detector models. However, calculations involving particle detectors are most commonly carried out in perturbation theory. Indeed, not only the justification of the model is jeopardized for strong couplings, but also some of the most interesting phenomenology, such emission and absorption of particles, can be described  at  quadratic order in the coupling strengths. Not only that, this is also the leading order for most phenomena in relativistic quantum information (e.g., detector's responses~\cite{Takagi}, communication~\cite{PhysRevD.101.036014}, entanglement harvesting~\cite{PhysRevD.94.064074} and the Fermi Problem~\cite{RevModPhys.4.87,PhysRevA.81.012330,PhysRevA.79.012304,PhysRevLett.107.150402}, etc..).   This section is devoted to analyze the order in perturbation theory at which superluminal propagation of initial data, described in last section, plays a role in measurement schemes involving more than two detectors. 
 
 Let us slightly extend the set-up described in section \ref{sec:propagation} by assuming that the unitary $\hat{U}$ in \eqref{unita} is implemented by a weakly coupled detector C, in such a way that we can write $\hat{U}=\hat S_{\textsc{c}}$. We can now determine at which order in perturbation theory the dynamics exhibits superluminal signalling, that is, at which order in perturbation theory condition \eqref{thirdcond} fails to hold.
 
 In order to do so, we first define the operator
   \begin{align}\label{fourthcond}
     \hat K\coloneqq[{\hat S}^{\dagger}_{\textsc{b}}\hat D_\textsc{b}{\hat S}_{\textsc{b}},{\hat S}_{\textsc{a}}\hat{S}_{\textsc{c}}{\hat S}^{\dagger}_{\textsc{a}}].
 \end{align}
 If this operator vanished there would be no superluminal propagation of initial data. We can determine the first order in the coupling strengths at which $\hat K$ does not trivially vanish.
 
 We can expand $\hat{K}$ in the coupling strengths by writing
 $\hat K=\hat K^{(0)}+\hat K^{(1)}+...$, where each $\hat K^{(j)}$ contains integrals involving $j$ Hamiltonians. Each term $\hat K^{(j)}$ will contain contributions from orders in the coupling constants of detectors C+A+B in such a way that all the powers add up to $j$.  It is easy to see that 
 \begin{align}
     \hat K^{(0)}=[\hat D_\textsc{b},\openone]=0,
 \end{align}
and that the linear term will also vanish
\begin{align}\label{k1}
\nn&    \hat K^{(1)}=\left[ \frac{\ii}{\hbar}\int_{-\infty}^{\infty} \d {t}
[\hat H_{\textsc{b}}({t}),\hat D_\textsc{b}],\openone\right]\\
\nn& +[ \hat D_\textsc{b},-\frac{\ii}{\hbar}\int_{-\infty}^{\infty} \d {t}
[\hat H_{\textsc{a}}({t}),\openone]]\\
&+[\hat D_\textsc{b} ,-\frac{\ii}{\hbar}\int_{-\infty}^{\infty} \d {t}
\hat H_{\textsc{c}}({t})]=0.
\end{align}
   The fact that the first two terms in \eqref{k1} vanish is obvious, while the third vanishes because B and C are spacelike separated.
   
   The higher order terms can be calculated similarly, but given the increasing complexity of the calculations it is more practical to reason which terms will vanish based on the following observations:
   \begin{enumerate}
       \item The zeroth order in the coupling constant of detector C cannot contribute to any order in $\hat{K}$, because at that order condition \eqref{thirdcond} is satisfied trivially.
       \item  The zeroth order in A cannot contribute at any order either, because B and C are spacelike separated.
       \item   Finally, the zeroth order in B cannot contribute at any order because in that case the induced observable ${\hat S}^{\dagger}_{\textsc{b}}\hat D_\textsc{b}{\hat S}_{\textsc{b}}$ acts trivially over the field.
   \end{enumerate}

  Therefore, $\hat{K}$  cannot have any contributions at quadratic order, because any quadratic contribution  will involve the zeroth order of at least one of the detectors. Hence,
  \begin{align}
  \hat K^{(2)}=0.
\end{align}

   This is not surprising if we take into account the result of section \ref{sec:faster}, because at quadratic order the detectors interact only binary. Since the detectors only influence each other pair-wise, the measurements cannot exhibit this type of superluminal signalling that involves necessarily three detectors. It is expected that this argument carries through in general calculations involving quadratic orders in perturbation theory.
   
   
   Interestingly, and perhaps less intuitively, the third order will also vanish. Indeed, the only term that can contribute at third order in perturbation theory, given the observations made above, is the one that involves the linear order of each in the three detectors.
   \begin{align}
\nn&    \hat K^{(3)}\\
& =-\frac{\ii}{\hbar^3}\left[ \int_{-\infty}^{\infty} \!\!\!\!\!\d {t}
[\hat H_{\textsc{b}}({t}),\hat D_\textsc{b}],[\int_{-\infty}^{\infty} \!\!\!\!\!\d {t}
\hat H_{\textsc{a}}({t}),
\int_{-\infty}^{\infty}\!\!\!\!\! \d {t}
\hat H_{\textsc{c}}({t})]\right].
\end{align}
   
   The operator in the first entry of the nested commutator acts over the space of detector B and over the quantum field, whereas the operator on the second entry is given by
   \begin{align}
     \int\d V\int \d V'[\hat {h}_\textsc{a}(\mathsf{x}),\hat{h}_{\textsc{c}}(\mathsf{y})].
   \end{align}
   Following the reasoning of section \ref{sec:propagation}, the microcausality condition forces this operator to be localized in the causal future of C, and therefore commutes with field operators localized in region B. We then conclude that
     \begin{align}
  \hat K^{(3)}=0.
\end{align}
   The interpretation in this case is that the superluminal propagation of initial data happens only at quadratic order in the detector that may act as a repeater. Since having a quadratic contribution in one of the detectors implies that at least one of the others contributes at zeroth order, the arguments given above force $\hat K^{(3)}=0$, and why no superluminal propagation can happen.

   The moral is that, assuming that all the measurements are weakly performed with detectors, impossible measurements are not present in most calculations done in the literature. One should be careful, however, when handling non-pertubative methods for smeared detectors.

\section{Conclusions}\label{sec:conclusions}
    
In this work we have presented a detailed account of the problem of formulating local measurements in QFT. We have explained how this problem motivates, both from the theoretical and practical points of view, the introduction of non-relativistic detector models. We have analyzed whether generalized Unruh-DeWitt-type detector models fulfill minimum requirements regarding relativistic causality. In other words, we have discussed whether non-relativistic systems coupled to quantum fields can be used to model repeatable  measurements on quantum fields without incurring in incompatibilities with relativistic causality. 

In particular,  we have investigated compatibility with relativistic causality in detector-based measurements by demanding that the signals emitted by each of the detectors should be constrained to lie within their associated future light-cones. Furthermore, we have formulated Sorkin's ``impossible measurements'' problem in terms of particle detector-based measurements, linking in this context the ``impossible measurements'' issues to the non-relativistic dynamics of the detector. The physical intuition is that, when a detector is spatially extended, the information propagating inside the detector is not constrained to travel subluminally since the detector is a non-relativistic system.  However, we have shown that within the usual assumptions---that is, weak couplings or point-like or nearly point-like detectors--- detector-based measurements are safe from the ``impossible measurement'' problem.

\acknowledgments

  The authors would like to thank Jason Pye for  discussions that were as long as they were helpful and insightful as well as Jos\'e Polo-G\'omez, Cendikiawan Suryaatmadja and the whole PSI RQI class of 2021 for finding elusive hidden typos in the darkest caverns of this manuscript. EMM acknowledges support through the Discovery Grant Program of the Natural Sciences and Engineering Research Council of Canada (NSERC). EMM also acknowledges support of his Ontario Early Researcher award. MP acknowledges support of the 2020 Constantine and Patricia Mavroyannis Scholarship Award by the AHEPA Foundation.

\appendix 
\section{Causal factorization in detector models}\label{app:factorization}
In this appendix we provide a proof of the causal factorization of the scattering operator for compactly supported detectors. The proof relies in elementary properties of unitary propagators. In this appendix we consider $\hbar=1$ in order to ease the notation.

Consider a general time-dependent interaction Hamiltonian of the form,
\begin{align}
    \hat H({t})= \hat H_{\textsc{a}}({t})+\hat H_{\textsc{b}}({t}),
\end{align}
and its associated Schrodinger equation
\begin{align}\label{esrrodinguer}
    \partial_{{t}}{\ket{\psi({t})}}_{\textsc{a}+\textsc{b}}=-\ii(\hat H_{\textsc{a}}({t})+\hat H_{\textsc{b}}({t}))\ket{\psi({t})},
\end{align}
or, more conveniently, in its integral form
\begin{align}
    \nn{\ket{\psi({t})}}_{\textsc{a}+\textsc{b}}&={\ket{\psi({t}')}}_{\textsc{a}+\textsc{b}}\\
    &-\ii\int^{{t}}_{{t}'}\d{t}(\hat H_{\textsc{a}}({t}'')+\hat H_{\textsc{b}}({t}''))\ket{\psi({t}'')}_{\textsc{a}+\textsc{b}}.
\end{align}
By recursively applying this integral equation, disregarding domain issues,  we can formally write the evolution of the state as the action of a two-parametric group of unitary operators, also known as the unitary propagator, $\hat U_{\textsc{a}+\textsc{b}}({t},{t}')$:
\begin{align}
   \nn&{\ket{\psi({t})}}= \hat U_{\textsc{a}+\textsc{b}}({t},{t}')\ket{\psi({t}')}\\
  &\nn =\sum_{n}\frac{(-\ii)^n}{n!} \int_{{t}'}^{{t}}\dots\int_{{t}'}^{{t}} \d {t}^{n}\\
&  \times \mathcal{T}(\hat H_{\textsc{a}}({t}_1)+\hat H_{\textsc{b}}({t}_1)\dots\hat H_{\textsc{a}}({t}_n)+\hat H_{\textsc{b}}({t}_n))\ket{\psi({t}')},
\end{align}
where the second line is the so-called Dyson expansion of the operator $\hat U_{\textsc{a}+\textsc{b}}({t},{t}')$.
Here we have defined the time ordering of two time-dependent operators $\hat A({t})$ and $\hat B({t})$ as
\begin{align}
   \nn& \mathcal{T}\hat A({t})\hat B({t}')\\
    &\coloneqq\theta({t}-{t}')\hat A({t})\hat B({t}')+\theta({t}'-{t})\hat B({t}')\hat A({t}),
\end{align}
where the definition is similar for higher orders.  It will be useful in the following to define unitary propagators that woudl be associated to local evolution, that is
\begin{align}
   \nn& \hat U_{\nu}({t},{t}')\ket{\psi({t}')}=\sum_{n}\frac{(-\ii)^n}{n!} \int_{{t}'}^{{t}}\dots\int_{{t}'}^{{t}} \d {t}^{n}\\
&  \times \mathcal{T}\hat H_{\nu}({t}_1)\dots\hat H_{\nu}({t}_n)\ket{\psi({t}')},
\end{align}
where $\nu\in\{\text{A},\text{B}\}$. In order to describe the dynamics of the detection process, we are particularly interested in the scattering operator, that is, the limit
\begin{align}
    \hat S_{\textsc{a}+\textsc{b}}=\lim_{{t}'\to-\infty} \lim_{{t}\to\infty}\hat U_{\textsc{a}+\textsc{b}}({t},{t}')
\end{align}
when the Hamiltonians are given by the expressions
\begin{align}
    \hat H_{\textsc{a},\textsc{b}}({t})=\int_{\mathcal{E}({t})}\!\!\!\!\!\d\mathcal{E}\;   \hat h_{\textsc{a},\textsc{b}}(\mathsf x).
\end{align}

Consider now that the supports of $\Lambda_{\textsc{a}}$ and $\Lambda_{\textsc{b}}$ are causally orderable, with A preceding B respect to some foliation (possibly different from $\mathscr{T}(\mathsf{x})$). Then the scattering matrix factorizes, i.e.
\begin{align}
    \hat S_{\textsc{a}+\textsc{b}}=\hat S_{\textsc{b}}\hat S_{\textsc{a}}.
\end{align}
To show this we find the Schrodinger equation of the factorized dynamics and prove that it coincides with the full dynamics. Then we will use a uniqueness argument to prove that therefore the dynamics coincide.

Consider the family of states 
\begin{align}
    {\ket{\psi({t})}}_{\textsc{a}\textsc{b}}=\hat U_{\textsc{b}}({t},-\infty)\hat U_{\textsc{a}}({t},-\infty) {\ket{\psi}_0}
\end{align}
where $\ket{\psi}_0$ is a fixed vector. It holds that
\begin{align}
    \nn&\partial_{{t}}{\ket{\psi({t})}}_{\textsc{a}\textsc{b}}\\
   &=-\ii\left(\hat H_{\textsc{b}}({t})+\hat U_{\textsc{b}}({t},-\infty)\hat H_{\textsc{a}}({t})\hat U^{\dagger}_{\textsc{b}}({t},-\infty)\right)\ket{\psi({t})}_{\textsc{a}\textsc{b}}.
\end{align}

Let us first distinguish two trivial cases.
First, consider that A precedes B, respect to the foliation $\mathscr{T}(\mathsf{x})$. Then, there exists a number 
${t}_c$ such that 
\begin{align}
    \hat H_{\textsc{b}}({t}_c)= \hat H_{\textsc{a}}({t}_c)=0
\end{align}
and
\begin{align}
    \hat H_{\textsc{b}}({t})=0\qquad {t}<{t}_c\\
    \hat H_{\textsc{a}}({t})=0\qquad {t}>{t}_c.
\end{align}
This  implies that
\begin{align}
    \hat U_{\textsc{b}}({t},-\infty)\hat H_{\textsc{a}}({t})\hat U^{\dagger}_{\textsc{b}}({t},-\infty)=\hat H_{\textsc{a}}({t})
\end{align}
for all ${t}$, since $\hat U_{\textsc{a}}({t},-\infty)$ is only different from the identity operator, $\hat\openone$, when $\hat H_{\textsc{b}}({t})=0$.

Second, if the supports are spacelike separated, then
\begin{align}
     [\hat H_{\textsc{b}}({t}),\hat H_{\textsc{a}}({t}')]=0
\end{align} 
for all ${t},t'\in\mathbb{R}$, and therefore
\begin{align}
 \nn & \hat U_{\textsc{b}}({t},-\infty)\hat H_{\textsc{a}}({t})\hat U^{\dagger}_{\textsc{b}}({t},-\infty)\\
  \nn &=\sum_{n}\frac{(-\ii)^n}{n!} \int_{-\infty}^{{t}}\dots\int_{-\infty}^{{t}} \d {t}^{n}\\
&  \times \mathcal{T}\left[\dots[\hat H_{\textsc{a}}({t}),\hat H_{\textsc{b}}({t}_1)]\dots,\hat H_{\textsc{b}}({t}_n)\right]=\hat H_{\textsc{a}}({t}).
\end{align}

More generally, assume that the detectors are causally orderable. Then, essentially (We are assuming, without proof, that the adjoint action of the unitary evolution can be carried inside the integral \eqref{ojocuidao}) we can write 
\begin{align}\label{ojocuidao}
 \nn & \hat U_{\textsc{b}}({t},-\infty)\hat H_{\textsc{a}}({t})\hat U^{\dagger}_{\textsc{b}}({t},-\infty)\\
&=\int_{\mathcal{E}({t})}\!\!\!\!\!\d\mathcal{E}\;\hat U_{\textsc{b}}({t},-\infty)\hat h_{\textsc{a}}(\mathsf x)\hat U^{\dagger}_{\textsc{b}}({t},-\infty).
\end{align}

For each $\mathsf{x}\in\text{supp}(\Lambda_\textsc{a})$, $\hat h_{\textsc{a}}(\mathsf x)$ is either causally connected or spacelike separated to the support of B,  so we can choose the corresponding proof from the two ones given above to show that it remains unchanged under the adjoint action of $\hat U_{\textsc{b}}({t},-\infty)$. Therefore 
\begin{align}
 \nn & \hat U_{\textsc{b}}({t},-\infty)\hat H_{\textsc{a}}({t})\hat U^{\dagger}_{\textsc{b}}({t},-\infty)\\
&=\int_{\mathcal{E}({t})}\!\!\!\!\!\d\mathcal{E}\; \hat h_{\textsc{a}}(\mathsf x)=\hat H_{\textsc{a}}({t}).
\end{align}

Altogether, the conclusion is that if A precedes B for some observer then $\ket{\psi({t})}_{\textsc{a}\textsc{b}}$ fulfils \eqref{esrrodinguer}, and since \eqref{esrrodinguer} is a linear differential equation, the vector
\begin{align}
    \ket{\varphi({t})}= \ket{\psi({t})}_{\textsc{a}\textsc{b}}-\ket{\psi({t})}_{\textsc{a}+\textsc{b}}
\end{align}
also fulfils \eqref{esrrodinguer}. Now, setting $ \ket{\varphi(-\infty)}=0$ implies $ \ket{\varphi({t})}=0$ for all ${t}$, since the solution is unique and $ \ket{\varphi({t})}=0$ is a solution with initial condition $\ket{\varphi(-\infty)}=0$. 

Therefore, we have shown that 
\begin{align}
    \ket{\psi({t})}_{\textsc{a}+\textsc{b}}= {\ket{\psi({t})}}_{\textsc{a}\textsc{b}}=\hat U_{\textsc{a}}({t},-\infty)\hat U_{\textsc{b}}({t},-\infty) {\ket{\psi}_0},
\end{align}
for all ${t}$, and more concretely,
\begin{align}
   \hat S_{\textsc{a}+\textsc{b}} {\ket{\psi}}= \ket{\psi(\infty)}_{\textsc{a}+\textsc{b}}\big |_{\ket{\psi(-\infty)}_{\textsc{a}+\textsc{b}}=\ket{\psi}}=\hat S_{\textsc{a}}\hat S_{\textsc{b}} {\ket{\psi}},
\end{align}
for all states in the Hilbert space.

\newpage
\bibliography{refs.bib}

\end{document}